\documentclass[12pt]{iopart}
\usepackage{color}
\usepackage{graphicx}
\usepackage[centerlast,margin=10pt,font=small,labelfont=bf]{caption}
\begin{document}

\newcommand{\expect}[1]{\langle #1 \rangle}
\newcommand{\creatop}[1]{\hat{a}^{\dagger}_{#1}}
\newcommand{\anniop}[1]{\hat{a}_{#1}}
\newcommand{\mi}{\mathrm{i}}
\newcommand{\mycomment}[1]{\textbf{#1}}
\newcommand{\bra}[1]{\langle #1|}
\newcommand{\ket}[1]{|#1\rangle}
\newcommand{\braket}[2]{\langle #1|#2\rangle}

\title{Simplified Quantum Process Tomography}

\author{M. P. A. Branderhorst, J. Nunn, I. A. Walmsley}
\address{Clarendon Laboratory, University of Oxford, Oxford OX1 3PU, UK}
%\affiliation{Clarendon Laboratory, University of Oxford, Oxford OX1 3PU, UK}
%\author{M P A Branderhorst}
%\affiliation{Clarendon Laboratory, University of Oxford, Oxford OX1 3PU, UK}
\author{R. L. Kosut}
\address{SC Solutions, 1261 Oakmead Parkway, Sunnyvale, CA 94085,USA}
%\author{I A Walmsley}
%\affiliation{Clarendon Laboratory, University of Oxford, Oxford OX1 3PU, UK}

\pacs{03.65.Wj}

\begin{abstract}
We propose and evaluate experimentally an approach to quantum process tomography that completely removes the scaling problem plaguing the standard approach. The key to this simplification is the incorporation of prior knowledge of the class of physical interactions involved in  generating the dynamics, which reduces the problem to one of parameter estimation. This allows part of the problem to be tackled using efficient convex methods, which, when coupled with a constraint on some parameters allows globally optimal estimates for the Kraus operators to be determined from experimental data. Parameterising the maps provides further advantages: it  allows the incorporation of mixed states of the environment as well as some initial correlation between the system and environment, both of which are common physical situations following excitation of the system away from thermal equilibrium. Although the approach is not universal, in cases where it is valid it returns a complete set of positive maps for the dynamical evolution of a quantum system at all times. 

\end{abstract}
\maketitle

\section{Introduction}
Quantum process tomography (QPT) provides a means to specify the complete map of a set of input states of a quantum system (say the system state at time $t=0$) to a set of output states (say the system state at some later time $t>0$). It is an essential tool for characterizing the dynamics of quantum systems, and is especially useful for systems undergoing non-unitary evolution \cite{nielsenQPT,nielsench}. Therefore it is useful not only for understanding the evolution of quantum system coupled to possibly unknown environments, but also for applications in which quantum systems are manipulated for particular ends. In practical realisations of quantum information processing, for example, quantum process tomography is necessary to fully characterize the operation of quantum logic gates, which is, in turn, critical in improving the fidelity of quantum computing devices. 
 
The main challenge in implementing QPT is the large number of required measurements. The positive map that represents the quantum process is specified by $N^{4}-N^{2}$ parameters for a system Hilbert space of dimension $N$. The number of experiments and the computational power required to estimate the process therefore scales exponentially with the size of the system specified in qubits. This makes is difficult to realise in systems of even modest dimension. 

A common experimental strategy to realise QPT involves the preparation of a complete set of input states that span the Hilbert space of the system. These input states act as probes of the quantum channel described by the positive map. After passing through the channel, the output states are reconstructed by means of quantum state tomography. The channel process is estimated by inversion of the information contained in the difference between the input and the output states.  Quantum process tomography has been implemented in optical systems \cite{opt_qpt1, opt_qpt2}, atoms in optical lattices \cite{steinbergQPT}, NMR \cite{kampermannQPT, nmr2} and a solid state qubit \cite{wrachtrup}.

A number of approaches exist that may reduce the number of input states that must needs be prepared. For example,  ancilla-assisted process tomography \cite{altepeter_qpt, darino_qpt} and direct characterization of quantum dynamics \cite{massoud_qpt} both use fewer probes than the direct approach, but the probes need to be entangled, which may be either resource-expensive or impossible in some experimental situations. Individual diagonal elements of a process matrix can be estimated efficiently from the average fidelities of appropriately modified channels, while off-diagonal elements require ancillas and controlled quantum gates \cite{Bendersky:2008kx,Bendersky:2009kx}.

In this article we present an approach to reduce the size of the quantum process tomography problem, based on convex optimization. A common assumption in quantum process tomography is that the process is completely unknown \textit{a priori}, it is a `black box'. The idea introduced in this article is based on the fact that in most cases in which quantum process tomography is wanted some knowledge will be available about what is going on inside the black box and that this prior knowledge could reduce the size of the problem significantly. One of the most elementary non-unitary processes is decoherence, caused by the coupling of a system to its environment. Decoherence manifests itself as phase- or amplitude damping. If it is known that decoherence is present, prior knowledge is available and the problem size can be reduced. The unknowns that need to be estimated are the coupling strength to the bath and the bath distribution function, neither of which can be measured directly. If a large enough and informative set of data is chosen these parameters can be estimated and not only the operator mapping the initial state onto a later one is known, but also the evolution at all times can be estimated. Thus from two sets of samples of the system state at two different moments in time the complete evolution can be inferred. We test this approach experimentally by characterizing the rotational dephasing of a vibrational wavepacket in diatomic potassium molecules. In this case the system size is more than two qubits, so that the feature of problem size reduction is emphasized. This is to our knowledge the first implementation of process tomography in molecules. 

%The total environment-system density matrix evolves unitarilyand at the second window the environment is traced out in the measurement, so the final system state $\rho_{\textnormal{\tiny{s,f}}}$ can be written as
%\begin{equation}\rho_{\textnormal{\tiny{s,f}}} = Q(\rho_{\textnormal{\tiny{s,i}}}) =  \mathrm{Tr_{\textnormal{\tiny{env}}}}(\hat{U}(\rho_{\textnormal{\tiny{s,i}}}\otimes\rho_{\textnormal{\tiny{env,i}}})\hat{U}^{\dag})\ \label{eq:krausmap}.\end{equation}
%Introducing the orthonormal environment basis $|k\rangle$ (USE MULTIMODE BASIS?) the output state can be written as
%\begin{equation}\rho_{\textnormal{\tiny{s,f}}} =\sum_{k}^{\kappa}\sum_{k'}^{\kappa}K_{kk'}\rho_{\textnormal{\tiny{s,i}}} K_{kk'}^{*}\ ,\end{equation}with completeness relation
%\begin{equation}\sum_{k}^{\kappa}\sum_{k'}^{\kappa}K_{kk'}^{*}K_{kk'} = I_{n}\ .\end{equation}

\section{Estimation procedure: incorporating prior knowledge}
In order to illustrate the new features of this approach,  we begin by developing the general theory of process tomography including an initially mixed quantum state of the enviroment and incorporating some prior knowledge of the Hamiltonian. We consider the unitary dynamics of the combined state $\rho_\mathrm{se}$ comprising the system and its environment,
\begin{equation}
\label{unitary}
\rho_\mathrm{se}(t)=U(t)\rho_\mathrm{se}(0)U^\dagger(t),
\end{equation}
where $U(t) = e^{-\mi H t}$ propagates the state over a time $t$ according to the time-independent Hamiltonian $H$. We suppose that initially, the system and environment are not correlated, so that $\rho_\mathrm{se}(0)=\rho_\mathrm{s}(0)\otimes \rho_\mathrm{e}(0)$. Over time, environmental interactions decohere the system, and we examine this decoherence by considering the dynamics of the reduced density matrix $\rho_\mathrm{s}$ for the system
\begin{equation}
\label{reduced}
\rho_\mathrm{s}(t) = \mathrm{Tr}_\mathrm{e}\left\{U(t)\rho_\mathrm{s}(0)\otimes \rho_\mathrm{e}(0)U^\dagger(t)\right\},
\end{equation}
where $\mathrm{Tr}_\mathrm{e}$ indicates a partial trace over the environment. This partial trace is conveniently performed in the basis $\ket{j}$ that diagonalizes the environment, taken to have dimension $J$. In this basis, the initial state of the environment can be written as a thermal ensemble
\begin{equation}
\label{init}
\rho_\mathrm{e}(0) = \sum_j^{J}  p_j \ket{j}\bra{j},
\end{equation}
and the system's dynamics reduce to the form
\begin{equation}
\label{dyn}
\rho_\mathrm{s}(t) = \sum_{jk} p_jE_{jk}(t)\rho_\mathrm{s}(0)E^\dagger_{jk}(t),
\end{equation}
where the Kraus operators $E_{jk}$ are given by
\begin{equation}
\label{kraus}
E_{jk}(t) = \bra{k}U(t)\ket{j}.
\end{equation}
The problem at hand is to estimate the probability distribution $\{p_j\}$ associated with the environment, as well as the Kraus operators themselves. 

This expression for the quantum process is known as the operator-sum representation, and is based on the set of Kraus operators $E=\{E_{kk'}|k,k'=1,...,J\}$. The basis can be transformed into a basis for which $J \leq n^{2}$ for an $n\times n$ dimensional system density matrix. This makes the physical interpretation of the Kraus matrices more difficult, although it simplifies the mathematics. The Kraus operators are not, of course, dependent on the experimental configuration, assuming that the environment itself is not under the control of the experimenter. Nonetheless, knowledge of the Kraus operators implies complete knowledge of the decoherence process. 

QPT is a means to estimate the Kraus operators from experimental data. A common procedure is to formulate the estimation algorithm as a convex optimization, that is, a search over all operators satisfying both the experimental data constraints and the mathematical form constraints of the problem in such a way as to guarantee that the solutions are globally optimal. Further, such problems may make use of efficient algorithms to identify such optimal solutions. Unfortunately, the estimation of Kraus operators in the form above is not a convex optimization problem, for two reasons. First, the equality constraint $\sum_{k}^{\kappa}E_{jk}^{\dagger}E_{jk} = I_{n}$ is not linear in $E_{jk}$ and second the objective function itself is not convex. Fortunately, the  problem can be reformulated into a convex form by expanding the Kraus operators in a fixed basis of system space operators.  Let $\{B_{i}\}$ be a set of $n^2$ operators that span the system Hilbert space. The Kraus operators can be expressed as a convex sum of these operators as \begin{equation}E_{jk} = \sum_{i=1}^{n^{2}}a_{ki}^{j}B_{i},\ k=1,...,J\ .\end{equation}
%For a mechanical oscillator, for example, a straightforward choice of basis is the vibrational eigenstates. 
Estimating the Kraus operators is then reduced to the estimation of a superoperator $X$, the representation of which is the fixed operators basis is given in terms of the matrix elements 
 \begin{equation}X_{\mu\nu} = \sum_{k=1}^{\kappa} a^{j*}_{k\mu}a_{k\nu}^{j},\  \mu,\nu = 1,...,n^{2},\
\end{equation}
where the superoperator is restricted to be positive and trace-preserving. Estimation of the superoperator from the data set is a convex optimization problem \cite{robert}.

The problem has now been made convex, but at the expense of an increase of the number of free variables. Since the size of the superoperator is  $n^{2}\times n^{2}$, and the size of the constraints is $n^{2}$, the total number of free variables is $n^{4}-n^{2}$. This number needs to be matched by the same number of variables in the measurements to estimate the process. This method is `expensive', even for experiments in which the collection of large data sets is straightforward.

Adoption of a few physically reasonable assumptions about the system-environment coupling, applied to the form of the Hamiltonian, greatly simplifies this task. In general we have
\begin{equation}
\label{ham}
H = H_\mathrm{s} + H_\mathrm{e} + H_\mathrm{se},
\end{equation}
where the first two terms represent the Hamiltonians of the system and environment, and where $H_\mathrm{se}$ is that describing their interaction. Two particularly simple special cases arise if (\emph{i}) $[H_\mathrm{s},H_\mathrm{se}]=0$, or (\emph{ii}) $[H_\mathrm{e} + H_\mathrm{se},\rho_\mathrm{e}(0)]=0$.

In the first case, the interaction Hamiltonian $H_\mathrm{se}$ can be decomposed in the same eigenbasis as the system Hamiltonian $H_\mathrm{s}$, so that no transitions between the system's energy eigenstates are induced by the coupling to the environment. The decoherence caused by the environment can then be described as `pure dephasing', and the Kraus operators can be written in the form
\begin{equation}
\label{dephasing}
E_{jk} = \sum_n e^{-\mi \omega_n t}\mu^{n}_{jk}(t)\ket{n}\bra{n},
\end{equation}
where the relation $H_\mathrm{s}\ket{n}=\hbar \omega_n\ket{n}$ defines the eigenstates and eigenfrequencies of the system, and where the time-dependent coefficients $\mu^n_{jk}$ are given by
\begin{equation}
\label{mus}
\mu^n_{jk}(t) = \bra{n}\otimes \bra{k} e^{-\mi(H_\mathrm{e}+ H_\mathrm{se})t} \ket{j}\otimes \ket{n}.
\end{equation}

 In the second case, the dynamics of the environment are diagonal in the same basis $\{\ket{j}\}$ that diagonalizes the initial state of the environment, meaning that different micro-states of the initial thermal ensemble are not coupled. The Kraus operators then take the form
\begin{equation}
\label{second}
E_{jk}=\delta_{jk}e^{-\mi K_j t}e^{-\mi H_\mathrm{s}t},
\end{equation}
where the $K_j$ are operators given by $K_j=\bra{j}(H_\mathrm{e}+H_\mathrm{se})\ket{j}$.

In the present work, we are fortunate that \emph{both} of the above situations obtain; case (\emph{ii}) exactly, and case (\emph{i}) approximately. The resulting expression for the system dynamics can be written in the form
\begin{equation}
\label{map}
\rho_\mathrm{s}(t)=\sum_jp_j \sum_{nm}\bra{n}\rho_\mathrm{s}(0)\ket{m}e^{-\mi (\kappa_{jn}-\kappa_{jm} + \omega_n-\omega_m)t}\ket{n}\bra{m},
\end{equation}
where the $\kappa_{jn}$ are the system eigenvalues of the $K_j$,
\begin{equation}
\label{kappa}
K_j\ket{n}=\kappa_{jn}\ket{n}.
\end{equation}
By inspection, it is clear that (\ref{map}) describes decoherence, since the off-diagonal elements of $\rho_\mathrm{s}$ involve summations over exponential phase factors with incommensurate frequencies. In fact, it is straightforward to show that under conditions (\emph{i}) or (\emph{ii}) the system evolves into the state
\begin{eqnarray}
\label{discord}
\rho_\mathrm{se}(t) &=& U(t)\rho_\mathrm{se}(0)U^\dagger(t)\nonumber \\
 &=& U(t)\ \rho_\mathrm{s}(0)\otimes \rho_\mathrm{e}(0)U^\dagger(t)\nonumber \\
 &=& \sum_jp_j \rho_\mathrm{s}^{j}(t) \otimes |j \rangle \langle j|.
\end{eqnarray}
Because the final state is a convex sum over system-environment product states, it is not entangled. However, because the final system states are not orthogonal,
\begin{equation}
\label{non-orthogonal}
\mathrm{Tr}(\rho_\mathrm{s}^{j}(t)\rho_\mathrm{s}^{j'}(t)) \neq 0,
\end{equation}
it does possess nonclassical correlations in the form of non-zero discord. It is the existence of such correlations that causes the decoherence of the system itself.

\section{Application: molecular vibrational decoherence by a rotational bath}
A prototypical system-environment interaction that serves to test this method of simplified process tomography is that of vibrational-rotational coupling in a diatomic molecule. In this case, a vibrational wavepacket (that is, a coherent superposition of eigenstates of the internuclear potential energy in a particular electronic state) is gradually rendered incoherent by means of dephasing. This arises because of the coupling of the vibrational and rotational degrees of freedom of the molecule, via the moment of inertia. The moment of inertia changes dynamically as the molecule vibrates, leading to a change in the rotational frequency, which, in turn, modifies the vibrational frequency. The net effect on the vibrational wavepacket is to dephase the superposition, leading to a mixed vibrational state. The aim of QPT in this case is to determine by experiment the quantum process that characterizes this dephasing, leading to an estimation of the Kraus operators in the parametrized form described in the previous section. 

The experimental procedure consists of exciting a vibrational wavepacket in an excited electronic state of a homonuclear dimer and monitoring its evolution by measuring the time-and frequency resolved fluorescence. This enables quantum state tomography (QST) of the mode as it evolves, and, as we shall show, QST provides sufficient information to implement simplified quantum process tomography (SQPT). 

\begin{figure}[tbp]
\centering
\includegraphics[width=8cm]{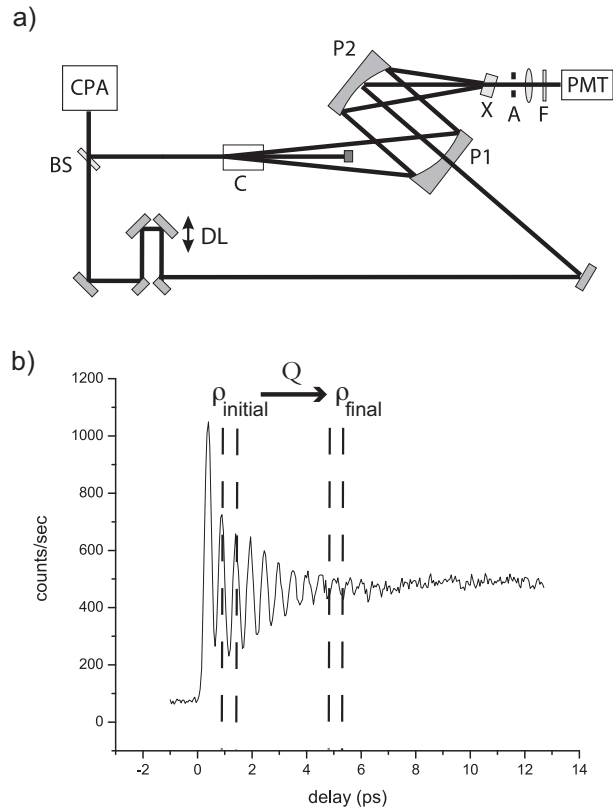}
\caption{(a) The experimental setup:  CPA -- laser system, BS -- beam splitter, DL -- variable delay line, C -- all-sapphire vapour cell, P$1$, P$2$ -- off-axis parabolic mirrors, X -- $3$ mm type I BBO crystal, A -- aperture and imaging lens, F -- IR-blocking passband filter, PMT -- photomultiplier; (b) the signal at the outer turning point, the data are collected within the two indicated time windows.} \label{fig:1}
\end{figure}

The experimental apparatus has been described in detail elsewhere \cite{Dunn93, Dunn95}  and is illustrated in figure \ref{fig:1}(a). Briefly, a vibrational wavepacket is prepared in the ${}^1\Sigma_{u}^{+}$ state of $\mathrm{K}_2$ through excitation by an ultrashort pulse, the vibrational period of the wavepacket is around 500\,fs. The molecules are kept in a vapour cell at $400\,{}^{\circ}\mathrm{C}$. At this temperature around 150 rotational levels are occupied, which creates a particularly detrimental environment for the vibrational mode, and causes rapid dephasing due to the modulation of the centrifugal coupling to the rotational degree of freedom. During the oscillation of the wavepacket in the excited state potential,  fluorescence is emitted when the electron makes a spontaneous transition to the ground state.  The fluorescence is imaged onto a nonlinear crystal, and, by mixing the fluorescence in the crystal with an ultrashort pulse so that it is gated in time and frequency, a tomographically complete data set for the vibrational mode can be collected, as described in detail in reference \cite{ref:oed}. 

From the time-frequency fluorescence map, we reconstruct the initial vibrational quantum state and the state at some later time. In order to do this we assume that there is a separation of timescales, such that data for QST is collected over a short enough period that there is no dephasing. Figure \ref{fig:1} (b) illustrates a typical fluorescence quantum beat pattern a particular wavelength, corresponding to the outer turning point of the vibrational wavepacket. The vertical dashed lines indicate the single vibrational periods that form the sampling windows for QST. The initial window occurs immediately following the termination of the exciting pulse, and the second at a sufficiently long time that significant dephasing has taken place. 

Quantum process tomography scales poorly with the size of the Hilbert space on which the process acts, and becomes quickly infeasible for larger Hilbert spaces. For the vibrational wavepacket discussed here, the Hilbert space dimension is $6$, so that the number of unknowns in the problem is $1260$. In order to implement "blind" quantum process tomography, the measurements should consist of $36$ orthogonal input states times 35 different settings, which we could do in principle by using differently shaped laser pulses. However, this is technically rather challenging, and becomes more so as we consider exciting more vibrational levels using broader bandwidth pulses.  The size of this testbed system provides a useful model by which to test ways in which the size of the problem may be reduced. In the particular case chosen here, a number of simplifications are possible based on knowledge of the Hamiltonian of the system and its environment. In fact, there are more than might normally be afforded. However, this serves to illustrate in a stark manner the extent to which some prior knowledge can render a significant gain in determining the quantum process. 

A first simplification is provided by the possible reduction of the number of input states. In standard process tomography experiments a complete set of input states is prepared, subjected to the process, and the output state is reconstructed. In this case the process is dephasing and a single input state is sufficient. Dephasing maps every element of the density matrix onto itself, so a single input state is sufficient as long as all elements of that density matrix are nonzero. 

The unitary evolution in equation %\ref{eq:krausmap} 
(\ref{unitary}) is determined by the Hamiltonian \begin{equation} 
\hat{H} = \frac{\hat{p}^{2}}{2\mu} +V( \hat{q}) + \frac{\hat{\textbf{J}}^{2}}{2 \mu
\hat{q}^{2}}\ ,\label{hamiltonian}
\end{equation}where $\hat{p}$ and $\hat{q}$ are the
internuclear momentum and position operators, $\mu$ is the reduced mass,
$V(\hat{q})$ is the adiabatic vibrational potential of the electronic ${}^1\Sigma_{u}^{+}$ state, and $\hat{\textbf{J}}$ is the angular momentum operator of the molecule orthogonal to the internuclear axis. The Hamiltonian can be simplified by approximating the electronic potential as being harmonic and the ro-vibrational centrifugal coupling term can be developed up to first order in the displacement parameter $\eta = \Delta q/ \bar{q}$ for small variations around the equilibrium internuclear separation $\bar{q}$ \cite{decoherence,decoherence2}. Writing the operator describing small displacements around this equilibrium position in terms of canonical creation and annihilation operators  $ \hat{a}^{\dag}$ and $\hat{a}$,\cite{decoherence} the Hamiltonian for a certain rotational level $j$ is 
\begin{equation}
\hat{H}_j = \hbar \omega \hat{a}^{\dag} \hat{a} + \hbar 6B\eta^2 j(j+1)\hat{a}^{\dag} \hat{a}+\hbar 3B\eta^2 j(j+1)(\hat{a}^{\dag 2} + \hat{a}^{2})\ ,
\end{equation} 
where $\omega$ is the vibrational frequency, $B$ the rotational constant. 
The Heisenberg equation of motion within a rotational subspace can be solved to yield the time dependent creation operator $\hat{a}_{j}^{\dag}(t) = \hat{a}^{\dag}(0)e^{i\omega t + i \lambda_{j}t}$ with $\lambda_{j}=6\eta^{2}Bj(j+1)$.  Now the evolution of state $|j,m\rangle |n\rangle$ within a $j$-subspace can be written as 
\begin{equation}
\hat{U}|j,m\rangle |n\rangle = \frac{\bigl( \hat{a}_{j}^{\dag}(t) \bigr)^{n}}{\sqrt{n!}}|j,m\rangle |0\rangle.
\end{equation} 
The Kraus operator arising from coupling to rotational bath mode $j$ is $\sum_{n}\langle j,m|U|j,m\rangle |n\rangle\langle n|$. Since the quantum number $m$ is degenerate we can sum over $m=-j,...,j$ which gives the Kraus operator
\begin{equation}E_{jk} = \delta_{jk}~N(2j+1)\sum_{n}e^{in\omega t + in\lambda_{j}t}|n\rangle\langle n|\ ,\end{equation} 
with $N$ the normalization factor. The state of the system at time $t$ is 
\begin{equation}
\label{evolved}
\rho_{\textnormal{\tiny{s,f}}} =  \rho_{\textnormal{\tiny{s}}}(t) =  \sum_{j}p_{j}E_{j}\rho_{\textnormal{\tiny{s}}}(0) E_{j}^{\dag},\,
\end{equation} 
where $p_{j}$ is the probability of rotational state $j$ being occupied in thermal equilibrium. The estimation of the superoperator is now narrowed down to estimation of the thermal probability distribution and the coupling constants $\{ \lambda_{j} \}$. Since at $400^{\circ}$C around $150$ levels are occupied, there are now only $300$ free variables, namely the set $\{ p_{j}, \lambda_{j} \}$. The number of variables may be reduced further by making use of the explicit form of the interaction Hamiltonian, in which $ \lambda_{j} = \lambda j(j+1)$, such that only a single coupling constant $\lambda$ is required. This is a large reduction of the required number of data required to characterize a completely unknown process, and therefore enables a larger system subspace to be taken into consideration. 

We solve the estimation of the distribution $\{ p_{j}, \lambda \}$ as a weighted least-squares problem with uniform weights. That is, a set of measurements is made to estimate the set of probabilities  $p_{\alpha\gamma} = \mathrm{Tr}(O_{\alpha\gamma} \rho_{s}(t))$, where $O_{\alpha\gamma}$ is the POVM representing the measurement, the experimental configuration of which is labelled $\gamma$, and  $\alpha$ is the measurement outcome. In our case $\gamma$ represents the settings that enable detection of fluorescence at a particular wavelengths and time delay after the excitation of the vibrational wavepacket, i.e. $\lambda = (\omega_{fluor}, \tau)$. The measurement outcome $\alpha$ is the strength of the fluorescence signal at that wavelength and delay. The experimentally determined probabilities is the set $p_{\alpha\gamma}^{\textnormal{\tiny{emp}}}$.

Then convex optimisation problem becomes
\begin{eqnarray}
\label{Minimize}
\mathrm{minimize}&&L(p_{j},D)=\sum_{\alpha\gamma}  \bigl(p_{\alpha\gamma}^{\textnormal{\tiny{emp}}} - \mathrm{Tr}O_{\alpha \gamma}\rho_{\textnormal{\tiny{s,f}}}\bigr)^{2} \\
\mathrm{subject\ to}~&& p_{j}\geq 0\ ,\
\sum_{j}p_{j}=1\ .
\end{eqnarray} 

The procedure is to measure the initial quantum state, and propagate it using the parametrized form of the Kraus operators. The expected probabilities $p_{\alpha\gamma}$ are subtracted from those empirically determined at the final time window $p_{\alpha\gamma}^{\textnormal{\tiny{emp}}}$, and the difference minimized by adjusting the parameters of the Kraus operators.  

The estimation of the coupling coefficients $\lambda_j$ is not a convex problem, since it is a Hamiltonian parameter \cite{robert}. Therefore we first estimate the optimal distribution of the bath distribution function for a fixed value of $\lambda$ as a semidefinite program using available interior-point method solvers, such as YALMIP.\cite{YALMIP} This is repeated for a number of values of the coupling parameter that are between those physically accessible in the experiment. These values are set by the inverse of the minimum temporal resolution of the delay and the range of delays. The minimum of the objective function $L$ over these two optimizations is taken as the global optimal, since the variation over $\lambda$ is shown to be convex across the feasible range. 

% around the physical value $6\eta^2Bj(j+1)$ to assess the trade-off between the coupling parameter and the bath distribution function. 

\section{Experimental results}

\begin{figure}[tbp]
 \centering
\includegraphics[width=11cm]{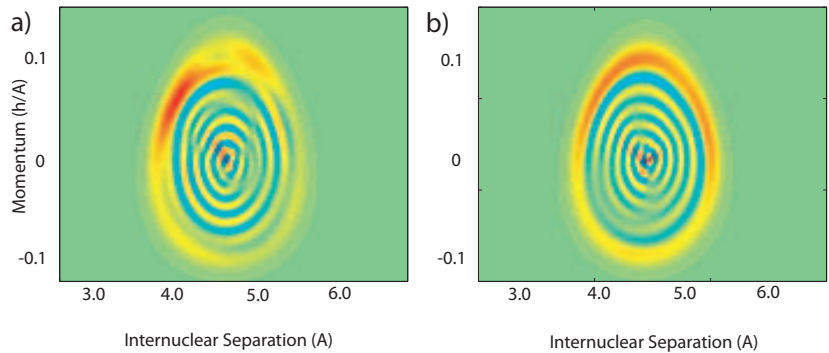}
\caption{Examples of the reconstructed Wigner distribution: (a) in the first window directly after excitation, and (b) in the second window, five periods later, after the decoherence process has acted. Note the latter distribution is more dispersed around the classical trajectory of the oscillator.}\label{fig:2}
\end{figure}

The quantum state of the system $\rho_{\textnormal{\tiny{s}}}$ is reconstructed by means of quantum state tomography \cite{ref:oed} based on an optimal experiment design protocol, using 250 homogeneously distributed experimental settings $\gamma$.  The initial ($t=0$) and final ($t=7$ ps) quantum states are shown in their Wigner representions in position-momentum phase space in figure \ref{fig:2}. The red and yellow regions show positive values of the Wigner function, whereas the darker, turquoise fringes are negative regions, indicative of non-classical character. 
The initially angularly localized Wigner distribution diffuses around the classical trajectory as a result of rotational dephasing. This corresponds to a decay of the off-diagonal elements in the density matrix. The purity of the initial state is $0.4$ and that of the final state is $0.2$, which is still big enough to observe quantum beats in the signal, since the corresponding Wigner distribution is not yet spread out over the entire trajectory. The inner product between the estimated initial density matrix $\rho_\mathrm{est,i}$ and the estimated final density matrix $\rho_\mathrm{est,f}$ is $\mathrm{Tr}(\rho_\mathrm{est,i}\rho_\mathrm{est,f}) = 0.15$, which provides a simple quantification of the dephasing process. 
Since the state tomography problem scales as $N^2$, for this case the number of unknown variables in the state tomography problem is $36$. 

%The number of unknown variables in the estimation of the bath distribution function is 150. The number of unknown variables in state tomography exceeds 150 if more than 12 vibrational levels are occupied. Also in that case the improvement is still significant, from $n^4-n^2$ to $n^2$.

\begin{figure*}[!hbt]
\centering
\includegraphics[width=10cm]{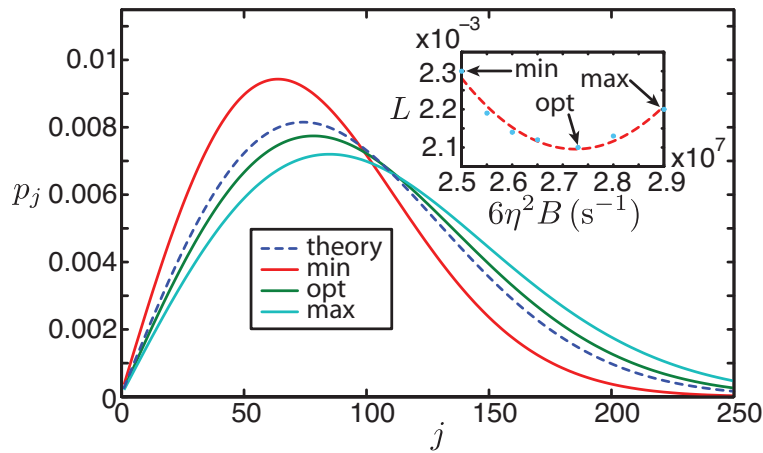}
\caption{The optimal distribution of weights $p_j$ over bath modes $j$ for three different values of the coupling coefficient: first the physical value $6\eta^2B=2.73\times10^7$ s$^{-1}$ (green), $2.5\times10^7$ s$^{-1}$ (red) and $2.9\times10^7$ s$^{-1}$ (cyan). The theoretical distribution is also plotted (dashed). Inset: The objective $L(p_j, D)$ evaluated as a function of the coupling parameter. Within the physically reasonable range of coupling parameters, the function is convex.} \label{fig:3}
\end{figure*}

The outcome of the estimation is shown in figure \ref{fig:3}. The bath distribution function has been estimated for several different values of the coupling parameter. The values span the physically feasible range, from $6\eta^2B=2.1\times10^7$ s$^{-1}$ to $2.9\times10^7$ s$^{-1}$, including the expected value, based on spectroscopic data, of $2.73\times10^7$ s$^{-1}$. Over this range, the objective $L(p_j, D)$ is convex, with a minimum value of $0.0021$, as shown in the inset. At the minimum of this curve, the bath distribution function is very  similar in shape to the rotational thermal distribution. The thermal bath distribution function is shown as the dashed line in figure \ref{fig:3}. The optimal distribution is plotted in green, and the distributions for the extremal values of the coupling parameter as red and cyan. The peak of the optimal distribution is close to that of the rotational distribution at the experimental temperature of $400$ K.

%\begin{figure*}[!hbt]
%\centering
%\includegraphics[width=14cm]{figure_temp}
%\caption{The thermal bath distribution function, similar to graph 1 of the estimated bath distribution function.} \label{fig:3}
%\end{figure*}

On one hand, the outcome is not surprising, because the protocol returns the thermal rotational distribution expected for a system in equilibrium. On the other hand, the outcome is non-trivial, because there is no direct access to the bath and the bath distribution function is estimated by measurements on the system alone.

The values of the objective $L$, as shown in the inset of figure~\ref{fig:3}, quantify the errors in the reconstructed measurement statistics. The differences between the empirical and estimated probabilities are so small that it is difficult to distinguish them by eye in a plot of the two sets of probabilities. We used a large data set for the estimation, and therefore the empirical probabilities are a good approximation to the true probabilities. These facts together suggest that our reconstructed statistics are a faithful rendering of the true probabilities.

\begin{figure*}[!hbt]
\centering
\includegraphics[width=14cm]{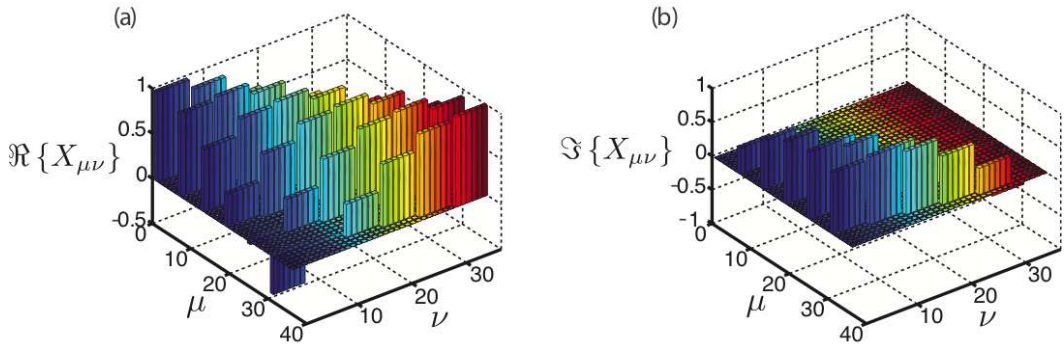}
%\caption{(a) The optimal distribution of weights $p_{j}$ over bath modes $j$. The outcome of the non-trivial estimation is similar to the the thermal distribution. (b) The real part of the $X$ matrix. The diagonal elements of the density matrix are not changed by the process; (c) the imaginary part of the $X$ matrix. The dephasing is larger for density matrix elements further away from the diagonal.} \label{fig:4}
\caption{(a) The real part of the $X$ matrix. The diagonal elements of the density matrix are not changed by the process. (b) The imaginary part of the $X$ matrix. The dephasing is larger for density matrix elements further away from the diagonal.} \label{fig:4}
\end{figure*}

Now it is possible to combine the estimated bath distribution $\left\{p_j\right\}$ and our prior knowledge --- encapsulated by the form of the map in Eq.~(\ref{evolved}) --- to construct the superoperator $X$. The real and imaginary part of the $36\times 36$ matrix are shown in figures \ref{fig:4} (b) and (c). The superoperator is positive semi-definite and trace preserving. Although the visual information from the figures is limited, one can recognize two general characteristics of the dephasing process: the process leaves the diagonal density matrix elements unchanged whereas the dephasing is larger for off-diagonal density matrix elements further away from the diagonal. 

To evaluate the precision of the current approach, we should ideally compare this estimated process with the true process. Of course we do not have access to the true process, but we can surmise that SQPT is reasonably accurate by comparing the final density matrix $\rho_\mathrm{s,f}$ --- as generated by propagating the initial reconstructed state $\rho_\mathrm{est,i}$ according to our reconstructed process --- with the final state $\rho_\mathrm{est,f}$ as reconstructed directly via quantum state tomography. We find the value of $\mathrm{Tr}\left\{\rho_\mathrm{est,f} \rho_\mathrm{s,f}\right\}$ to be $0.19$, which is close to the purity $\mathrm{Tr}\{\rho_\mathrm{est,f}^2\}=0.2$.

\section{Conclusions}
 
We have formulated and demonstrated a method to reconstruct completely the decoherence process that is caused by the coupling of a vibrational system to its rotational environment in a diatomic molecule. This method avoids the main problem of process tomography which is the scaling of the problem. The way in which our approach avoids this problem is to apply some prior knowledge of the quantum process. This information consists of an assumption about the form of the coupling of the system to its environment, and provides a general means for simplification. The particular conditions under which it can be applied are broadly applicable - the linearity of the system-environment coupling, and the commutativity of the bath or system operators with the coupling Hamiltonian - though they are not universal. The problem is thereby reduced to one of parameter estimation, in which the number of parameters is vastly smaller than the number of elements of the process operators. Further, it provides a route to including realistic properties of the environment, such as a non-zero temperature, as well as constructing the quantum channel for any particular time evolution. The problem of estimating the enviroment distribution function at finite temperature has the form of a convex optimization problem, which makes the problem easy to solve numerically, and guarantees the solution is the global optimum. Estimating the coupling parameters, which are part of the Hamiltonian, is not strictly convex, but there are few enough parameters to make searching the space straightforward and not computationally onerous. We anticipate that the formalism can, for example, be generalized for the case of a quantum logic gate in the presence of dephasing. and to other systems in which non-unitary dynamics plays a role. 

\section*{References}

%\bibliography{../../References/references}

\end{document}